 \documentclass[onecolumn]{aastex63}

\usepackage{amsmath}
\usepackage{bm}

\def\vec#1{\ensuremath{\bm{#1}}}

\received{--}
\revised{--}
\accepted{--}

\submitjournal{ApJ}

\shorttitle{Modelling of coronal transverse oscillations driven by p-modes}
\shortauthors{Gao et al.}

\usepackage{amsmath}
\usepackage{mathtools}
\usepackage{CJK}

\begin{document}

\title{Modelling of transverse oscillations driven by p-modes in short coronal loops}

\correspondingauthor{Mingzhe Guo}
\email{mingzhe.guo@kuleuven.be}

\begin{CJK*}{UTF8}{gbsn}
\author[0000-0002-6641-8034]{Yuhang Gao (高宇航)}
\affiliation{School of Earth and Space Sciences, Peking University, Beijing, 100871, People's Republic of China}
\affiliation{Centre for mathematical Plasma Astrophysics, Department of Mathematics, KU Leuven, Celestijnenlaan 200B bus 2400, B-3001 Leuven, Belgium}

\author[0000-0003-4956-6040]{Mingzhe Guo}
\affiliation{Centre for mathematical Plasma Astrophysics, Department of Mathematics, KU Leuven, Celestijnenlaan 200B bus 2400, B-3001 Leuven, Belgium}

\author[0000-0001-9628-4113]{Tom Van Doorsselaere}
\affiliation{Centre for mathematical Plasma Astrophysics, Department of Mathematics, KU Leuven, Celestijnenlaan 200B bus 2400, B-3001 Leuven, Belgium}

\author[0000-0002-1369-1758]{Hui Tian}
\affiliation{School of Earth and Space Sciences, Peking University, Beijing, 100871, People's Republic of China}
\affiliation{Key Laboratory of Solar Activity and Space Weather, National Space Science Center, Chinese Academy of Sciences, Beijing 100190, People's Republic of China}

\author[0000-0002-3814-4232]{Samuel J. Skirvin}
\affiliation{Centre for mathematical Plasma Astrophysics, Department of Mathematics, KU Leuven, Celestijnenlaan 200B bus 2400, B-3001 Leuven, Belgium}

\begin{abstract}

Recent observations have revealed two types of decayless transverse oscillations in short coronal loops: 
one with short periods scaling with loop lengths, 
and the other with longer periods that exhibit a peak at around 5 min in the period distribution. 
To understand such a difference in period, 
we work in the framework of ideal MHD and
model a short coronal loop embedded in
an atmosphere with density stratification from the chromosphere to the corona.
An inclined p-mode-like driver with a period of 5 min is launched at one loop footpoint. 
It is discovered that two types of decayless transverse oscillations 
can be excited in the loop.
We interpret the 5 min periodicity as being directly driven by the footpoint driver,
while the others, with periods of several tens of seconds, are regarded as kink eigenmodes of different harmonics. 
Therefore, our simulation shows that both types of decayless oscillations found in observations can be excited by p-modes in one short coronal loop.
This study extends our understanding of ubiquitous decayless transverse oscillations in the corona. 
Furthermore, it suggests that p-modes could be an important energy source for coronal heating by driving decayless transverse oscillations.

\end{abstract}

\keywords{Magnetohydrodynamics (1964); Solar atmosphere (1477); Solar oscillations (1515); Solar coronal loops (1485); Solar coronal waves (1995); Magnetohydrodynamical simulations (1966)}

\section{Introduction} \label{sec:intro}

Transverse oscillations in coronal loops are a subject of intensive investigations in solar physics since their first imaging observation in 1999 \citep{naka1999,aschwanden1999}. 
They can be used as a powerful diagnostic tool of the coronal magnetic field \citep[e.g.,][]{naka2001,verwichte2013,chen2015,su2018,yang2020b,yang2020a}, and can also contribute to our understanding of coronal heating \citep[see the review by][]{tvd2020}. 
External energy release processes can excite decaying kink oscillations in coronal loops \citep{zimovets2015,zhang2020,li2023sutri}.
These oscillations begin with a large amplitude, and then rapidly decay in several periods \citep[e.g.,][]{nistico2013,goddard2016}. 
Coronal loops can also support transverse oscillations without discernible excitation processes. 
Such oscillations were first discovered by \cite{wang2012} and \cite{tian2012} through imaging and spectroscopic observations respectively at nearly the same time. They have a smaller amplitude without any obvious damping, and are therefore called decayless oscillations \citep{nistico2013}. 
Both oscillations are believed to be standing kink waves, 
and their period linearly scales with the loop length in observations \citep[e.g.,][]{anfin2015,goddard2016}.

Decayless oscillations are found to be ubiquitous in active region loops \citep{anfin2015} and can be continually observed for a long time \citep[e.g.,][]{zhong2022long}. Previous observational studies mainly focused on oscillations of long coronal loops with a length of several hundred megameters, and used data from the Atmospheric Imaging Assembly \citep[AIA;][]{lemen2012} on board the Solar Dynamics Observatory \citep[SDO;][]{pesnell2012}. Their typical periods range from 1.5 to 10 min with an average displacement amplitude of $\sim0.17$ Mm \citep{anfin2015}.
Recently, decayless oscillations have been also found in short coronal loops ($\lesssim$ 50 Mm) with the high-resolution 174 $\mathring{\text{A}}$ images taken by the Extreme Ultraviolet Imager \citep[EUI;][]{rochus2020} on board the Solar Orbiter \citep[SolO;][]{muller2020}. 
\cite{Petrova2023} reported high-frequency oscillations in
two coronal loops with lengths of 4.5 Mm and 11 Mm located in a quiet-Sun region. The measured periods are 14 s and 30 s, much shorter than in longer coronal loops. 
\cite{zhong2022eui} also identified such oscillations with an average period of 1.6 min in a 51-Mm long coronal loop using simultaneous observations of EUI and AIA. Furthermore, \cite{li2023} conducted a statistical study of decayless short-period oscillations in 111 loops observed by EUI, and obtained a median period of 40 s and loop lengths of 10-30 Mm. The scaling law was found to be similar to that for longer loops, suggesting that all the oscillation events are of the same type, namely the (fundamental) kink eigenmode. Based on that, the authors estimated the magnetic field with the measured oscillation parameters following the seismology method used in previous studies \cite[e.g.,][]{naka2001,tian2012,nistico2013,anfi2019}.

Decayless transverse oscillations have also been detected in coronal bright points \citep[CBPs;][]{tian2012,gao2022}, 
which are generally composed of many small magnetic loops \citep[see the review of][]{madjarska2019}. \cite{gao2022} investigated 31 decayless oscillation events in CBPs using EUV imaging observations obtained with AIA. The oscillation period is 1-8 min while the loop length ranges from 14 to 42 Mm. Unlike previous observations \citep{anfin2015,Petrova2023,li2023}, no obvious linear correlation was found between the oscillation period and loop length. 
Meanwhile, the distribution of the period shows a peak at about 5 min \citep[see Figure 3 in][]{gao2022}. 
The most straightforward interpretation is that the transverse oscillations are driven by a periodic external driver like photospheric p-modes since their period is also around 5 min.
Note that Figure 6 in \cite{zhong2022eui} also reveals 
a similar peak in the histogram of periods. 

In fact, photospheric p-modes have been suggested to excite transverse waves and oscillations in the corona for a long time. Many observations show that an enhanced power occurs at $\sim$3 mHz in the coronal velocity power spectrum, implying that the observed transverse waves may be excited by p-modes \citep[e.g.,][]{tvd2008,tomczyk2009,morton2015,morton2016,morton2019}. 
Such a power spectrum has also been used to drive kink oscillations in simulated coronal loops \citep[e.g.,][]{pagano2019,howson2023}.
Meanwhile, wave excitation by p-modes is also widely studied in a number of analytical and modeling works. Some previous works show that transverse waves can be generated from p-modes through mode conversion \citep[e.g.,][]{khomenko2012,cally2017}, while \cite{riedl2019} and \cite{skirvin2023} have found another scenario that an inclined p-mode driver can directly excite kink oscillations in a flux tube.

In this study, we perform a 3D magneto-hydrodynamic (MHD) simulation to study the relationship between p-modes and the excitation of two types of decayless transverse oscillations observed recently in
short coronal loops. 
This paper is organised as follows: Section \ref{sec:methods} describes our model and numerical setup. 
The simulation results are presented in Section \ref{sec:results}.
In section \ref{sec:discussion}, we further discuss the results and relate them to observations. 
Finally, Section \ref{sec:summary} summarizes our findings.

\section{Model} \label{sec:methods}

\begin{figure*}
    \centering
    \includegraphics[width=1.0\textwidth]{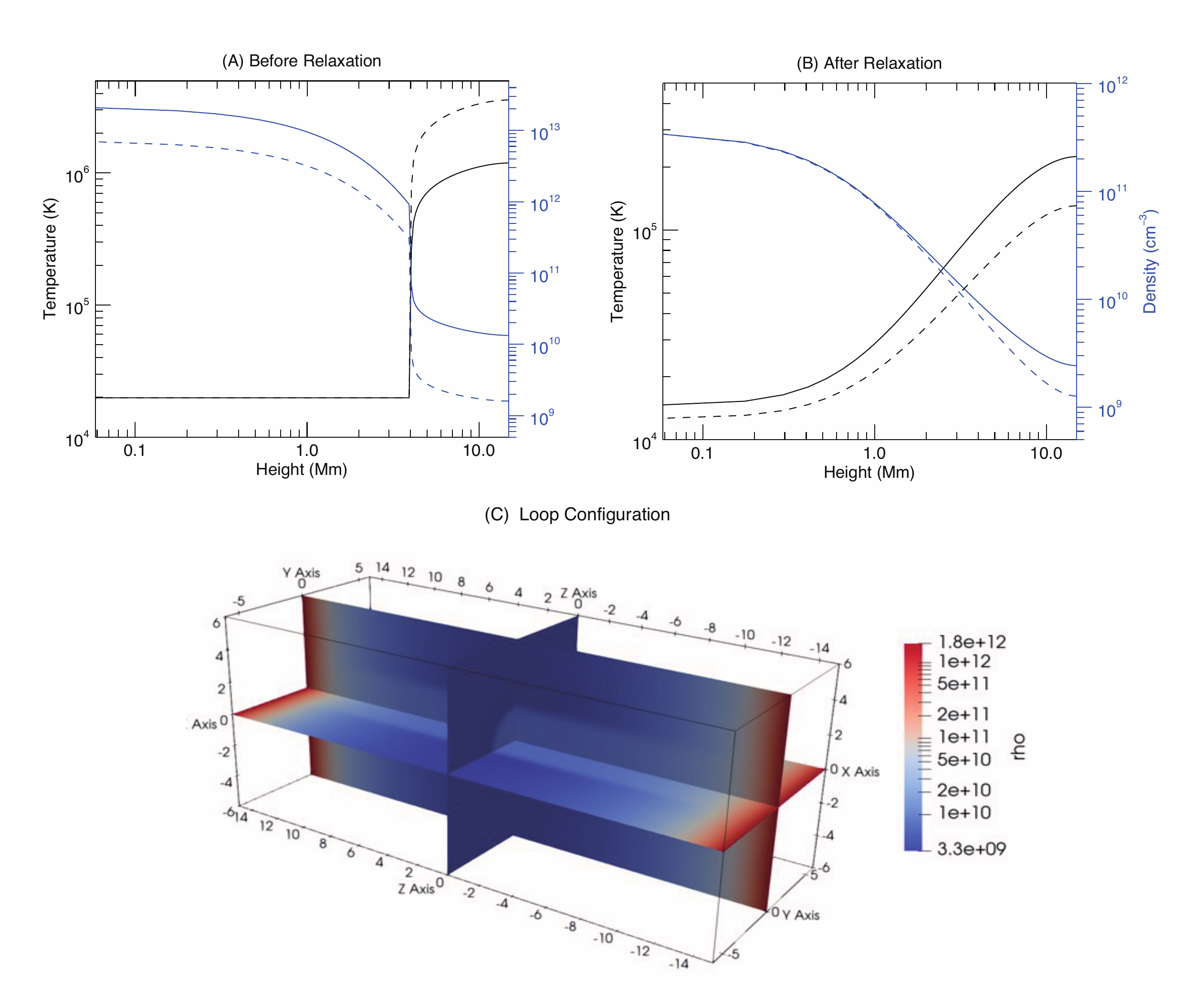}
    \caption{Top two panels present the profiles of temperature (black) and number density (blue) along the loop axis ($z$-direction) before (A) and after (B) the 2D relaxation. The horizontal axis ``Height" is actually the increase of $z$ with respect to the footpoint.
    The solid and dashed lines represent the value inside and outside the loop. Panel (C) shows the 3D loop configuration.}
    \label{fig:model}
\end{figure*}

In this study, we aim to model a short coronal loop with a length of 30 Mm, which is close to the average loop length found in \cite{gao2022}. 
Gravitational stratification and the lower atmosphere (i.e., chromosphere and transition region) are included \footnote{As a result, the coronal part of our loop is shorter than 30 Mm, as illustrated in Figure \ref{fig:model}(C).}.
Therefore, we employ a magnetic flux tube similar to that used in
\cite{pelouze2023} and \cite{guo2023}.
The simulation is divided into two steps:
a 2D run first working in cylindrical coordinates ($r,z$) and then a 3D run in Cartesian coordinates.
In both runs, 
the gravity is specified as $g(z)=g_\odot \sin(\pi z/L)$ along the loop, 
where $L=30$ Mm is the loop length, 
$g_\odot=274 $ m s$^{-2}$ is the gravity at the solar surface.
Such a distribution is widely used in previous studies \citep[e.g.,][]{kohutova2017,kara2019a,kara2020,soler2021,riedl2021,pelouze2023}. 
The temperature profile from the chromosphere to the corona is given by: 
\begin{equation}
    T(r,\,z)=\left\{\begin{aligned}
        &T_\text{ch},& z \le z_\text{ch},\\
        &T_\text{ch}+\left[T_\text{co}(r)-T_\text{ch}\right]\left[1-\left(\frac{L/2-z}{L/2-z_\text{ch}}\right)^2\right]^{0.3},& z> z_\text{ch},\\
    \end{aligned}\right.
\end{equation}
which is derived from \cite{aschwanden2002}. 
The temperature and height of the chromosphere are $T_\text{ch}=2\times 10^4$ K and $z_\text{ch}=4$ Mm, respectively. The chromosphere is set to be thicker than the real case, following some previous works \citep[e.g.,][]{howson2022,pelouze2023,guo2023}. Nevertheless, the physical properties of the chromosphere are appropriately described and the thick chromosphere can be regarded as an extended boundary. 

The coronal temperature $T_\text{co}$ has a transverse distribution:
\begin{equation}
    T_\text{co}(r)=T_\text{e}+\frac{1}{2}(T_\text{i}-T_\text{e})\left\{ 1-\tanh\left[\left(\frac{r}{R}-1\right)b\right] \right\}.
\end{equation}
Here the temperature inside and outside the loop is set to be $T_\text{i}=1.2$ MK and $T_\text{e}=3.6$ MK, respectively. 
$R$ represents the loop radius and $b$ is a dimensionless number determining the thickness of the inhomogeneous layer. 
Initially, they are set to be 1 Mm and 10 respectively, corresponding to an inhomogeneous layer of 0.6 Mm. Similarly, the density profile at the bottom boundary is set to be
\begin{equation}
    \rho_\text{ch}(r)=\rho_\text{e,ch}+\frac{1}{2}(\rho_\text{i,ch}-\rho_\text{e,ch})\left\{ 1-\tanh\left[\left(\frac{r}{R}-1\right)b\right] \right\}\,,
\end{equation}
where the internal (external) density is set to be $\rho_\text{i,ch}=3.51\times 10^{-8}\text{ kg m}^{-3}$ ($\rho_\text{e,ch}=1.17\times 10^{-8}\text{ kg m}^{-3}$). 
The density at larger heights is set to satisfy the field-aligned hydrostatic equilibrium, as described in \cite{pelouze2023}. The vertical profiles for temperature and density at the initial state are shown in Figure \ref{fig:model}(A).

A uniform magnetic field of 42 G along the $z$ axis is adopted in the whole simulation domain. Given that short coronal loops have a lower height and mainly distribute in active regions and CBPs, such a magnetic field strength can be reasonable \citep[e.g., see][]{Wang2007,tian2008,nistico2013,madjarska2019}. 

As indicated by \cite{pelouze2023} and \cite{guo2023}, such an initial setup is not in magnetohydrostatic (MHS) equilibrium. As a result of the non-uniform transverse structuring, obtaining an analytical pressure balance equilibrium is not possible. Therefore, we relax the system numerically until the background velocities are small enough to be neglected. 
To save computational time,
we first conduct the relaxation in the 2D simulation. 
To suppress the initial vertical flows in the loop, 
we multiply the three velocity components by a factor $\alpha(z,t)$ at each time step during the 2D simulation. 
Similar to \cite{pelouze2023}, the factor is defined as:
\begin{equation}
    \alpha(z,t)=\left\{
    \begin{aligned}
        & 1, & t\le t_{0}\text{ or } t>t_1,\\
        & 1-0.05\left(\frac{L/2-|z|}{L/2}\right)\left(\frac{t-t_{0}}{t_{0}}\right), & t_{0}<t\le 2t_{0},\\
        & 1-0.05\left(\frac{L/2-|z|}{L/2}\right),& 2t_{0}<t\le t_{1}.
    \end{aligned}
    \right.
\end{equation}
The velocity is multiplied by the factor $\alpha<1$ between time $t_0$ and $t_1$. 
In practice,
we choose $t_0=2 \tau$ and $t_1=4500\tau$, where $\tau=7.78$ s is the unit time. 
After $t_1$, we continue to conduct the relaxation for another 1500 $\tau$, and succeed in obtaining an equilibrium state. 
Figure \ref{fig:model}(B) illustrates temperature and density profiles along the loop axis after the relaxation process. The horizontal flows have been suppressed to a value below $0.01$ km s$^{-1}$ in the whole simulation domain, which indicates that a new MHS equilibrium has been achieved. The magnetic field inside the loop is now slightly different from that outside \citep[see also][]{pelouze2023}. Note that the transition region has been broadened to several megameters. It is a compromise hard to avoid since in reality, the transition region is very thin and contains very large density and temperature gradients. Considering this, artificially broadening the transition region has become a widely-used approach that allows a coarser resolution in the vertical direction \citep[e.g.,][]{lionello2009,mikic2013}.

We then rotate the 2D relaxed results to 3D Cartesian coordinates as the initial state for the next 3D MHD simulations, 
as shown in Figure \ref{fig:model}(C).
To ensure that there are no unphysical effects arising from the coordinate transition, 
we then let the 3D system evolve for another 300 $\tau$.
After that, an inclined p-mode-like driver is employed at one footpoint ($z=-15$Mm) to study waves excited in the loop. 
It is described by:
\begin{equation}\label{eq:driver}
    \vec{v}(x,y,t)=V_\text{0}\sin\left(\frac{2\pi t}{P}\right)\exp\left(-\frac{x^2+y^2}{\sigma^2}\right)\Vec{e}_\theta,\quad \vec{e}_\theta=(\sin\theta,0,\cos\theta).
\end{equation}
The driver has a velocity amplitude ($V_\text{0}$) of 500 m s$^{-1}$, 
a period ($P$) of 300 s, and an inclination angle ($\theta$) of 15$^\circ$ relative to the loop axis. It is expected that the inclination can excite transverse Alfv\'enic motions/oscillations in the loop by breaking the azimuthal symmetry of the system \citep[see][]{riedl2019,skirvin2023}. 
To ensure that the driver is located inside the loop, we multiply it with a Gaussian function with a form of $\exp(-\frac{x^2+y^2}{\sigma^2})$ and a standard deviation $\sigma=2$ Mm. 
Note that the loop radius has increased to about 2 Mm after relaxation.

For both the 2D and 3D runs,
we employ the PLUTO code \citep{mignone2007} to solve the ideal MHD equations.
For spatial reconstruction, a piece-wise TVD linear scheme is utilized, while numerical fluxes are computed using the Roe Riemann solver. A second-order characteristic tracing method is applied for time stepping. Anisotropic thermal conduction is also incorporated into our simulation while explicit resistivity and viscosity are not included.
To maintain the divergence-free nature of the magnetic field, 
we employ the hyperbolic divergence cleaning method \citep[e.g.,][]{dedner2002}. 
In the 2D case,
the computational domain is $[0, 6]$ Mm $\times [-15, 15]$ Mm, with a uniform grid of
$128\times 512$ cells, resulting in a resolution of 47 km in the $r$-direction, and 59 km in the $z$-direction.
In the 3D simulation, 
we use a uniform grid of 256 cells from $-6.0$ to $6.0$ Mm in both the $x$ and $ y$ directions, resulting in a resolution of 47 km.
In the $z$-direction,
128 grid cells are adopted from $[-7.5,7.5]$ Mm and
512 grid points are uniformly distributed in 
the left two domains near the loop ends,
namely $[-15.0,-7.5]$ Mm and $[7.5,15.0]$ Mm.
This ensures a higher resolution near the two footpoints to reveal the dynamics of the lower atmosphere.

The boundary conditions are described as follows. During the 2D relaxation process, we adopted an axisymmetric boundary condition at $r = 0$ (the loop axis) and an outflow boundary condition at $r=6$ Mm.
At two footpoints of the loop ($z = \pm 15$ Mm), we extrapolated the density and pressure from the hydrostatic equilibrium. 
Meanwhile, we followed the method used by \cite{kara2019a} to set the magnetic field, which can ensure that the normal gradient of magnetic field equals zero.
Additionally, the radial velocity $v_r$ and vertical velocity $v_z$ are set to be reflective.
For the 3D simulation, we used outflow conditions for all the lateral boundaries, and similar conditions as the 2D case for the vertical boundaries.
We only changed the velocity conditions according to Equation (\ref{eq:driver}) at one footpoint ($z=-15$ Mm) to mimic the p-mode driver.

\section{Results} \label{sec:results}

\begin{figure*}
    \centering
    \includegraphics[width=1.0\textwidth]{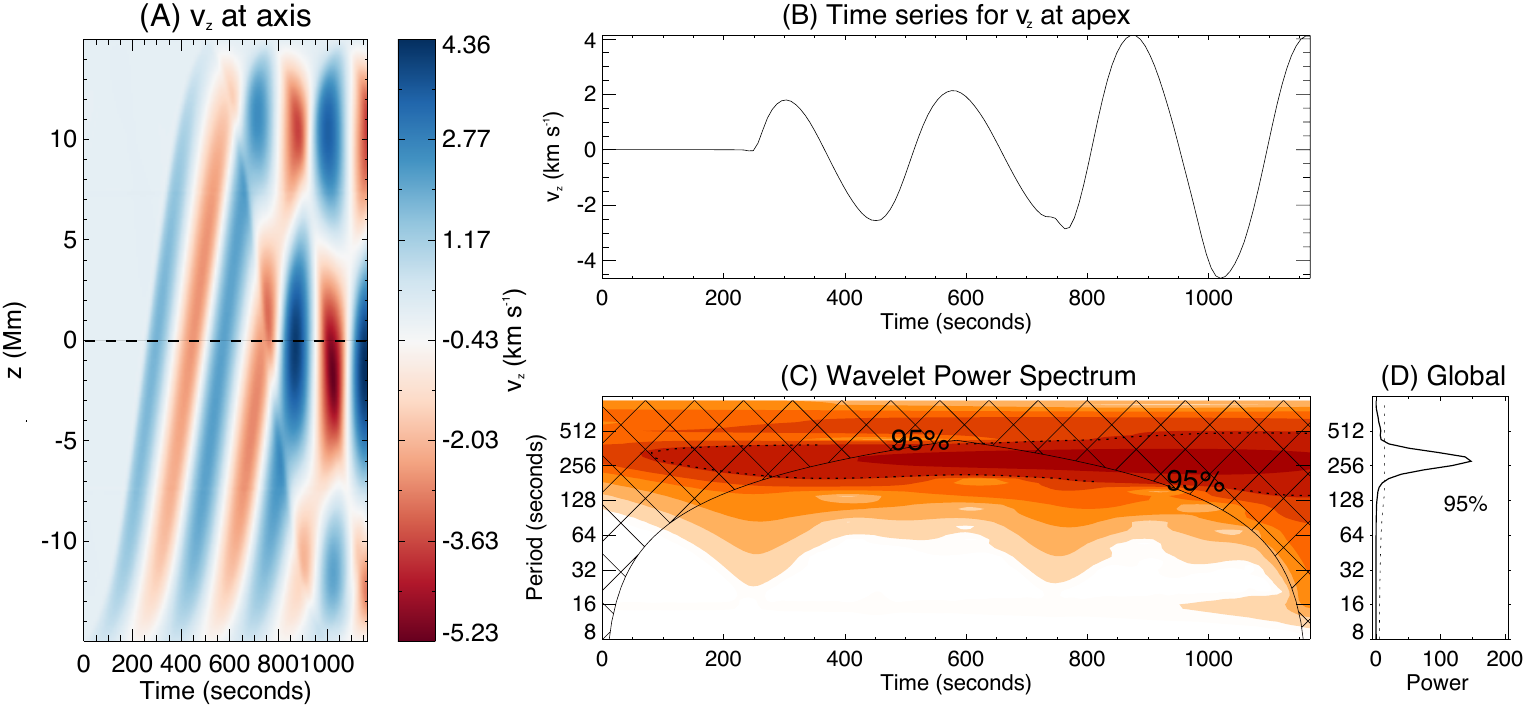}
    \caption{(A) Space-time plot for longitudinal velocity perturbation $v_z$ at the loop axis. (B) Time series of $v_z$ at the loop apex ($z=0$), corresponding to the dashed black line in (A). (C-D): Wavelet analysis results of the time series shown in (B). A darker color represents a larger wavelet power. The dashed lines denote a 95\% significance level.}
    \label{fig:vz}
\end{figure*}

In this section, 
we will present our numerical results. 
Note that a convergence study has been conducted to ensure that the 
current results are not influenced by the numerical resolution.
As shown in previous studies \citep{riedl2019,skirvin2023}, an inclined p-mode driver can excite longitudinal waves, 
especially slow sausage modes, in the loop. 
Figure \ref{fig:vz}
presents the evolution of vertical/longitudinal velocity $v_z$.
The $z-t$ distribution of $v_z$ is shown in Figure \ref{fig:vz}(A). 
Seen from the slope, the vertical velocity perturbation propagates along the loop with a speed of about 90 km s$^{-1}$ in the corona, which is close to the local sound speed $c_\text{s}$. 
The propagation is slower in the lower atmosphere, resulting from a smaller $c_\text{s}$ there (see also Figure \ref{fig:speeds}(B) and the relevant discussion in Section \ref{subsec:cutoff}). 
After about 500 s, the velocity distribution is affected by the reflection at another footpoint ($z=15$ Mm), forming a pattern like the third-harmonic oscillation at the end of the simulation. 
Figure \ref{fig:vz}(B)-(D) show the time series for $v_z$ at the loop apex ($z=0$) and the associated wavelet analysis results.
It takes about 240 s for the wave signal to reach the apex, then the plasma oscillates there with a period of $\sim$280 s, which is close to the period of the driver.

\begin{figure*}
    \centering
    \includegraphics[width=1.0\textwidth]{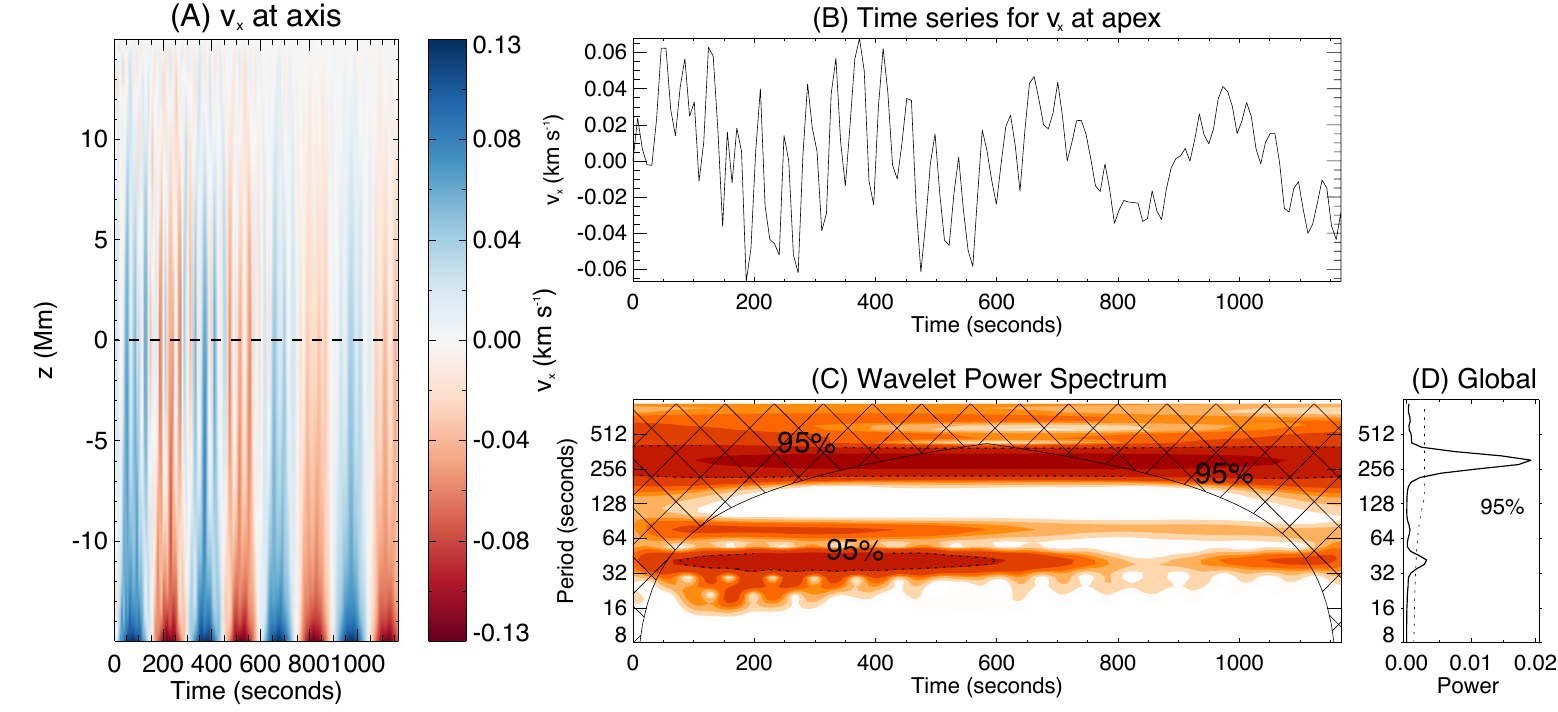}
    \caption{Similar to Figure \ref{fig:vz} but for the transverse velocity perturbation $v_x$.}
    \label{fig:vx}
\end{figure*}

In this study, we are more interested in the transverse waves/oscillations excited in the loop. 
As Figure \ref{fig:vx}(A) illustrates, transverse oscillations are quickly excited and can reach the loop apex with a much larger propagation speed than slow waves (see also Section \ref{subsec:cutoff}). 
It can also be seen in the time series of $v_x$ at the loop apex (Figure \ref{fig:vx}(B)), as the wave signals arrive there much earlier compared to the longitudinal motions as shown in Figure \ref{fig:vz}(B). Meanwhile, the oscillations display both long and short-period regimes. 
From the wavelet spectrum shown in Figure \ref{fig:vx}(C) and (D), we can find that there are actually three different periods, 305.2 s, 76.3 s and 41.6 s. 
The longest period is close to 5 min, namely, the period of the driver, which means that the long-period oscillation is directly driven by the p-modes similar to the longitudinal motions. Specifically, it can be regarded as a propagating kink wave with a large wavelength that is longer than the loop length.\footnote{Another possibility is to consider such oscillation as a quasi-static evolution in the loop in response to a slow oscillating driver, similar to the motion of a guitar string when one end is fixed and the other end is manually moved aside.}
As for the higher frequency oscillations, we suggest that they correspond to the eigenmode kink oscillations of the loop, where the two different periods detected here should be related to different harmonics. Given that the wavelet power is derived from the transverse velocity at the apex, we expect to find only odd harmonic signals here. As a result, the period of 76.3 s should be related to the fundamental mode, while 41.6 s likely correspond to the third harmonic. 
Interestingly, the global wavelet power at 41.6 s is much higher than that at 76.3 s, which is different from the expected case. This point will be further discussed in Section \ref{subsec:harmonic}. 

Besides the oscillation period, the transverse velocity amplitude at the loop apex is also investigated. The apparent amplitude of the long-period oscillation at the loop apex is about 0.06 km s$^{-1}$ at the beginning (before $t\sim 600$ s) and then decreases to about 0.04 km s$^{-1}$. From the wavelet power spectrum (Figure \ref{fig:vx}(C)), the 5-min oscillation has a nearly constant power, indicating a non-decaying nature. In fact, it is as expected since the footpoint driver persistently offers energy to this periodicity. Therefore, we suppose that, from the perspective of observation, such an oscillation would appear like the decayless oscillation, although it is actually a driven motion. 
On the other hand, the short-period oscillations seem to experience a damping with time, which could be a result of resonant absorption \citep[e.g.,][]{goossens2011}. However, as seen in panel (B), a high-frequency oscillation without apparent damping can exist for more than ten cycles. As a comparison, decaying oscillations always damp quickly in less than 6 cycles \citep[e.g.,][]{naka1999,goddard2016}.
Hence, it is reasonable to regard both high and low-frequency oscillations as decayless transverse oscillations in an observational sense.
However, the velocity amplitude in our simulation is smaller than previous observations \citep[e.g.,][]{gao2022}. This could be because the simulation driver is different from the real case. Considering a linear scaling between the amplitude of the excited oscillation and that of the driver \citep[see e.g.,][]{poedts1990,kara2019b}, we can expect that a larger $V_\text{0}$ or $\theta$ in Equation (\ref{eq:driver}) can increase the transverse amplitude in the loop. Also, in such a closed system, the pile-up of energy will gradually increase the amplitude \citep[see e.g.,][]{kara2019a,guo2019b}. In addition, another possible reason for the small amplitude could be related to the cut-off in the transition region, which will be discussed in detail in Section \ref{subsec:cutoff}.

\begin{figure*}
    \centering
    \includegraphics[width=1.0\textwidth]{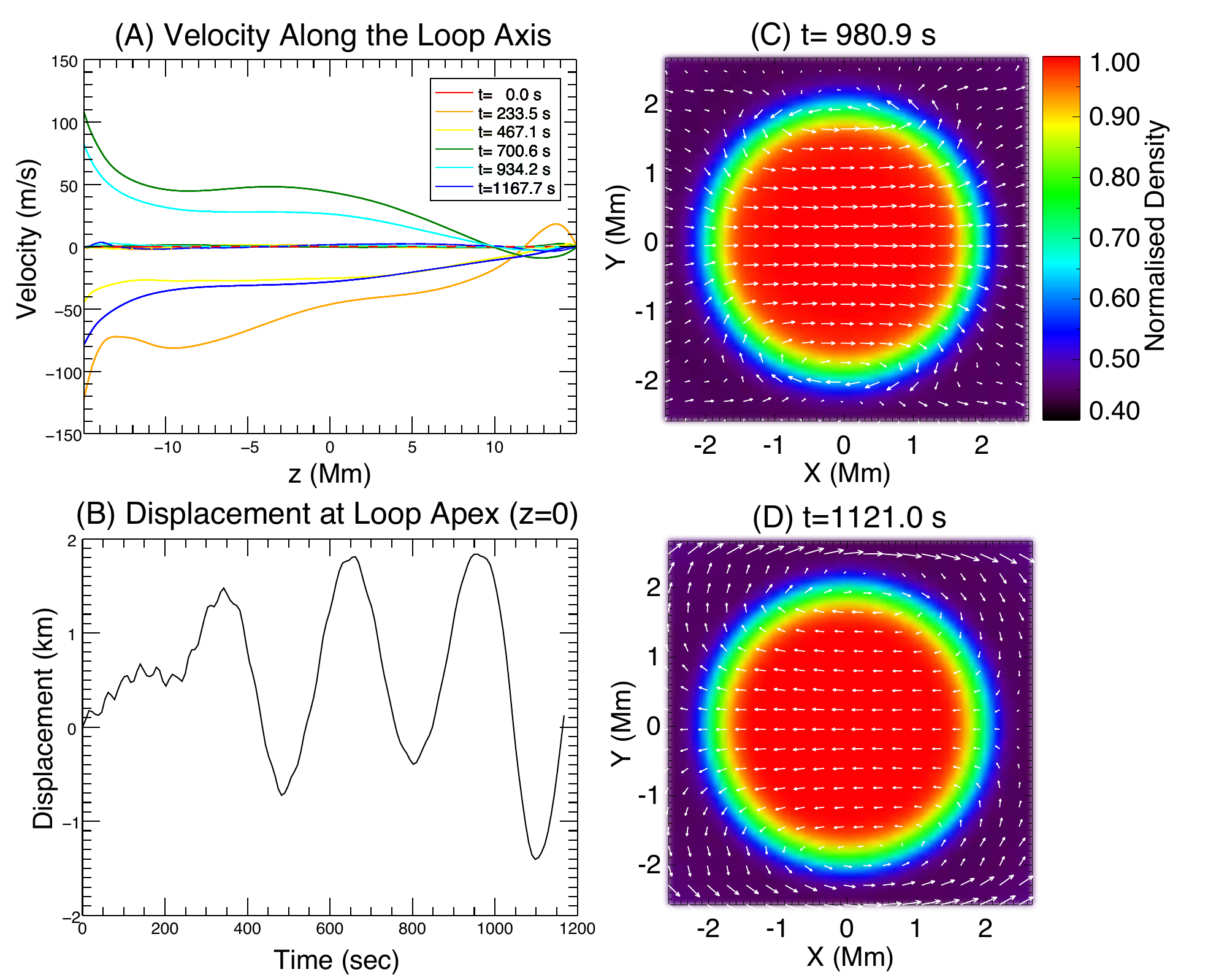}
    \caption{(A) Horizontal velocity along the loop axis at several time steps, with solid lines representing $v_x$ and dashed lines representing $v_y$ (multiplied by 10 because the value of $v_y$ is very low). (B) Transverse displacement of the loop obtained from considering the center of mass. (C)(D) The cross-section profiles of the normalised density at $z=0$ (loop apex) at $t=980.9$ s and $t=1121.0$ s, with the horizontal velocity marked by white arrows.}
    \label{fig:kink}
\end{figure*}

In Figure \ref{fig:vx}, we choose the $x$ component of velocity ($v_x$) to illustrate the transverse oscillations. However, we need to be careful because sausage mode or fluting mode is also accompanied by the fluctuation of horizontal (radial) velocity. Therefore, in order to prove the existence of transverse kink mode with an azimuthal wave number $m=1$, we need to further confirm whether the axisymmetry of the loop is broken \citep[see][]{skirvin2023}. In Figure \ref{fig:kink}(A), we plot $v_x$ and $v_y$ along the loop axis at several different time steps. We notice that $v_y$, the horizontal velocity component in the direction perpendicular to the driver, is much smaller than $v_x$, thus showing a non-axisymmetric property. We also determine the displacement of the loop by calculating the position of the center of mass. The time series of displacement at the loop apex ($z=0$) is shown in Figure \ref{fig:kink}(B). Roughly speaking, the oscillation of displacement has a similar pattern to that of $v_x$ in Figure \ref{fig:vx}(B), but contains less details. Nevertheless, the coexistence of two different period regimes can be also observed. In Figure \ref{fig:kink}(C) and (D), we plot two snapshots of the cross-section density profile (normalised) at the loop apex, with the horizontal velocity field overplotted. The distributions of the velocity vectors are consistent with the typical characteristics of the $m=1$ kink mode, i.e., the “dipole-like” pattern \citep[e.g., see][]{goossens2014,guo2020,skirvin2022}. Based on these results, we can conclude that there are dominant transverse kink oscillations excited in the loop.

\section{Discussion}\label{sec:discussion}



\subsection{Two types of decayless transverse oscillations in short coronal loops}\label{subsec:2types}

In our simulation, we found that two types of transverse oscillations are excited by a velocity driver that is reminiscent of an inclined p-mode. 
The first one has periods of several tens of seconds, suggesting that it is of the same type as the high-frequency oscillations found in recent EUI observations \citep{Petrova2023,li2023}. 
In other words, these oscillations are kink eigenmodes (this fact will be further verified in Section \ref{subsec:harmonic}), and the period is determined by the physical parameters of the loop. 
Specifically, the period is proportional to $L/c_\text{ph}$, where $c_\text{ph}$ is the phase speed. In the long-wavelength limit, $c_\text{ph}$ is approximately equal to the kink speed $c_\text{k}$ \citep{er1983}, which is further determined by the density and magnetic field inside and outside the loop. In the corona, the kink speed does not vary greatly with time and location, leading to a roughly linear scaling between $L$ and $P$ as observed in \cite{anfin2015,goddard2016} and \cite{li2023}. To the best of our knowledge, our simulation shows that such oscillations in short coronal loops could be excited by the p-modes for the first time.

The second type of oscillation excited in our simulation has a period of about 5 min. It can be seen as a propagating wave whose period is modulated by an external driver, namely, the p-mode-like continuous velocity perturbation we add at the footpoint. We suggest that such a scenario may explain the observational results of decayless oscillations of short coronal loops in CBPs \citep{gao2022}. Moreover, it is easy to understand why no linear correlation between the loop length (wavelength) and period was found in this observation, because the frequency is not determined by the loop length $L$. 
 
In fact, the decayless transverse oscillations found in both longer and shorter coronal loops could be a mixture of these two types of oscillations. \cite{zhong2022eui} summarised a number of previous observational results of oscillation periods and loop lengths \citep{wang2012,nistico2013,anfin2013,anfin2015,duckenfield2018,anfi2019,mandal2021,zhong2022long,Petrova2023}, and made a scatter plot between these two parameters, as well as a histogram of the period. There is not only a linear scaling in the scatter plot, but also an obvious peak at about 300 s in the period distribution. Additionally, the scatter relative to the best linear fit between $P$ and $L$ is the largest at around 4-5 min. This may imply that the 5-min transverse oscillations related to the p-mode leakage are likely to occur in those observational studies, although these oscillations were all considered as the kink eigenmode. Furthermore, we note that in reality, p-modes are known to be stochastically driven and have different periods, leading to a period distribution with a peak at $\sim5$ min, as illustrated in \cite{tomczyk2009} (see their Figure 2). In our model, we just adopt a mono-periodic driver with a 5 min period for simplicity. However, if we consider the fact that p-modes can have varying periods, we could have a broad-band period distribution of the driven transverse oscillations, which can be even closer to the observational results \citep[e.g.,][]{gao2022,zhong2022eui}.

In observations, the coexistence of two types of transverse oscillations in one oscillating loop as shown in Figure \ref{fig:vx} has not been found yet. 
It could be attributed to the limitation of instruments. 
Most of the previous observational studies analysed data from SDO/AIA \citep{wang2012,nistico2013,anfin2013,anfin2015,duckenfield2018,anfi2019,mandal2021,zhong2022long,gao2022}. However, AIA's temporal resolution (12 s) of EUV bands limits our ability to observe high-frequency oscillations. 
This could explain why \cite{gao2022} failed to detect such oscillations in CBPs. On the other hand, the data from High Resolution Imager (HRI) of EUI has both higher spatial and temporal resolution than AIA, allowing us to study short-period oscillations in shorter coronal loops \citep{Petrova2023,zhong2022eui,li2023}. 
But unfortunately, EUI/HRI has only a limited field of view, and the continuous observation time is only several hours at most.
Meanwhile, the shorter loops observed so far tend to be more dynamic, and most oscillation events just last for 2-4 cycles.
Therefore, it is difficult to detect signals of the 5-min oscillation. 
Nevertheless, future EUI observations could still hopefully capture some short coronal loops that can exist stably for a long duration, and thus the simultaneous existence of these two types of oscillations might be observed.

\subsection{Excitation mechanism of the decayless oscillation}\label{subsec:excitation}

So far, the excitation mechanism of ubiquitous decayless oscillations is still under debate. 
Unlike decaying oscillations which are excited by external pulses, 
decayless oscillations can persist for multiple cycles while retaining the oscillation amplitude.
Previous studies have confirmed mechanisms that are responsible for the damping of kink oscillations in the coronal loop, 
such as resonant absorption \citep[e.g.,][]{goossens2011,guo2020}, formation of Kelvin-Helmholtz instability \citep[e.g.,][]{heyvaerts1983,terradas2008}, and turbulence \citep{tvd2021}.
These mechanisms may also be present in decayless oscillations, despite their lower transverse amplitudes \citep[see e.g.,][]{antolin2016,kara2019a,kara2019b,guo2019a,guo2019b,shi2021}. 
In this case, however, 
a persistent external energy input is required to keep the amplitude from decaying. 
According to different types of drivers, current excitation models can be divided into three categories \citep[e.g., see][]{naka2021,gao2022review}: 1) a simple harmonic driver with a single frequency at the footpoint of the loop \citep[e.g.,][]{nistico2013,Yuan2023}. 2) an external quasi-steady flow such as the super-granulation motions \citep[also named the self-oscillatory model; see][]{naka2016,kara2020,kara2021}. 3) a random driver with a broad-band frequency \citep[e.g.,][]{afa2020}.

In this study, we use a model inspired by \cite{pelouze2023} and \cite{guo2023} but to study decayless oscillations driven by p-modes that are inclined to the vertical axis of the loop structure.
Previous 3D MHD simulations often added a transverse driver with a period matching the eigenfrequency \citep[e.g.,][]{kara2019a,kara2019b,guo2019a,guo2019b,guo2023}. 
However, our results show that a 5-min driver can also excite kink eigenmodes. 
To some extent, this mechanism may be similar to the self-oscillatory model, since the driver has a much larger period than that of the kink eigenmode in the short loop, and thus can be considered as a quasi-steady flow. Meanwhile, the driver can also excite another oscillation regime with a similar period, which distinguishes it from the typical self-oscillatory model. In any case, the oscillation-generating scenario in this study can be regarded as a new excitation mechanism for decayless oscillations, especially in short coronal loops, which is also consistent with recent observations. 
The link of widespread decayless oscillations to p-modes implies a potential energy source for wave heating of the corona.

\subsection{Harmonics of the eigenmode kink oscillation}\label{subsec:harmonic}

\begin{figure*}
    \centering
    \includegraphics[width=1.0\textwidth]{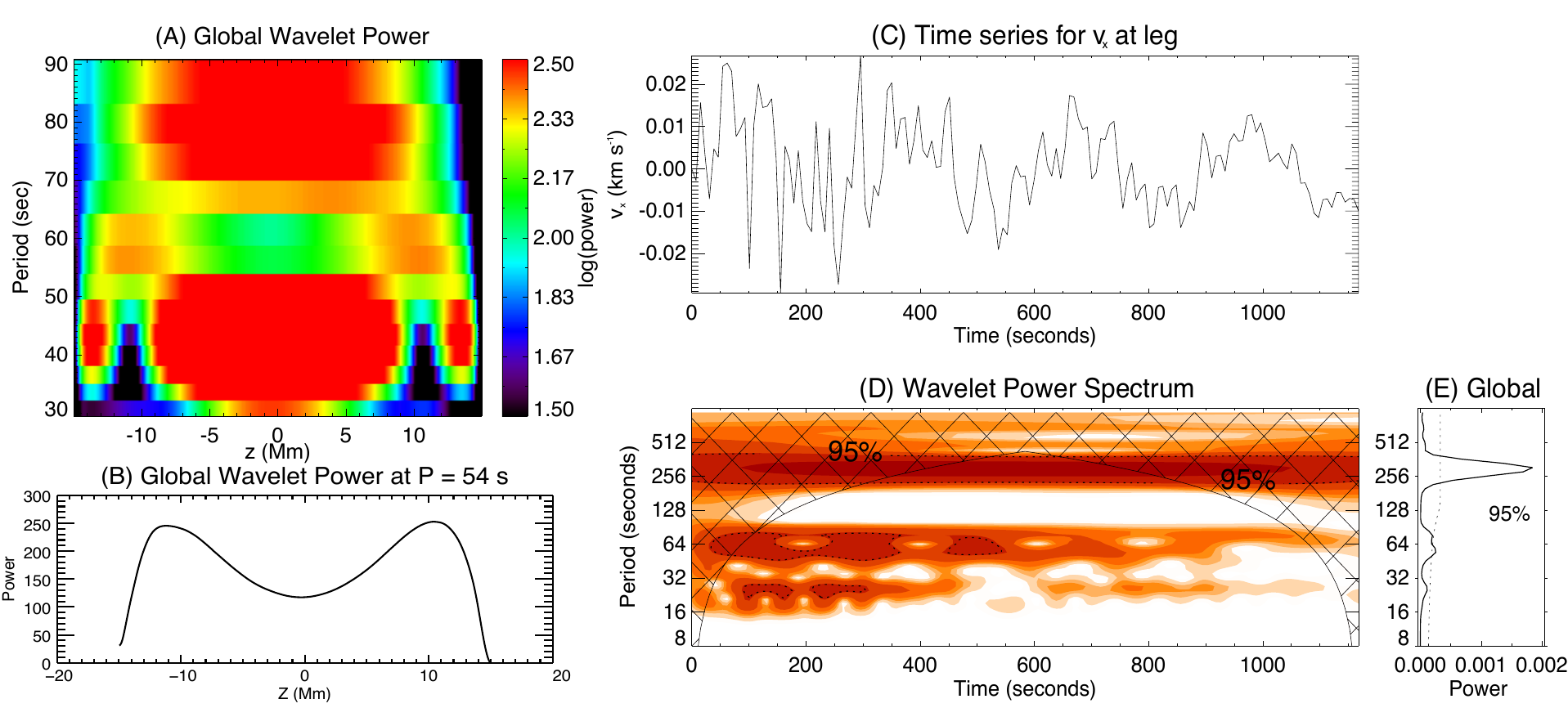}
    \caption{(A) The global wavelet power derived from time series of $v_x$ at different positions ($z$) along the loop axis. (B) The global wavelet power at the period of 54 s along the loop axis. (C-E) Similar to Figure \ref{fig:vx}(B-D) but for $v_x$ at the loop leg ($z=11$ Mm) rather than the loop apex. }
    \label{fig:harmonic}
\end{figure*}

Researchers are always interested in the harmonic characteristics of kink oscillations in coronal loops, since the period ratio can be used to diagnose the plasma structures in the corona \citep[e.g.,][]{andries2005,tvd2007,duckenfield2018}. Previous modelling works on decayless oscillations have revealed the excitation of high harmonics in loops \citep{afa2020,kara2020,kara2021}. In this study, wavelet analysis of $v_x$ at the loop apex clearly revealed the existence of different harmonics of the eigenmode kink oscillation (see Figure \ref{fig:vx}(C)), and they are characterised as the fundamental mode with a period of 76.3 s and the third harmonic with a period of 41.6 s.

Here, we display Figure \ref{fig:harmonic} to demonstrate the presence of the second harmonic. We calculate the global wavelet power spectrum at all positions along the loop axis, and the results are shown in Figure \ref{fig:harmonic}(A). 
To highlight the harmonic signals, we take the logarithm of the power. As for the vertical axis relating to the period, we only present a range of 30-90 s, corresponding to the range where the kink eigenmode periods lie. From the figure, we can see the second-harmonic oscillation with a period of 54.0 s has also been excited besides the fundamental mode and the third harmonic. The global wavelet power at 54.0 s for different $z$ is also shown in panel (B). The location of the peak power can be determined as $z=\pm 11$ Mm.  The time series for $v_x$ here are also analysed, and the results are shown in Figure \ref{fig:harmonic}(C-E). The wavelet analysis revealed a new peak at about 54.0 s corresponding to the second harmonic which is not present in Figure \ref{fig:vx}(D). 
In summary, we can conclude that there are at least three harmonics excited in our simulation.

Considering the density stratification and lower atmospheric parts in our model, it is difficult to analytically calculate the period of kink eigenmodes in the loop. Therefore, we choose to numerically determine this period by running our simulation with an initial velocity impulse transverse to the loop rather than a continuous footpoint driver, to excite kink eigenmodes \citep[see e.g.,][]{guo2023}. 
Specifically, we consider three different initial velocity perturbations. All of them are in the $x$-direction and restricted inside the loop, with an initial distribution as $v_0 \sin[\pi n\left(z/L+1/2\right)]$ where $v_0=5$ km s$^{-1}$.
The distribution has the form of standing waves with $n$ antinodes, and thus the velocity perturbations can excite transverse kink oscillations of the $n$-th harmonic. We conduct simulations for $n=$ 1, 2, and 3, and successfully obtain the eigenmode period for these three harmonics, i.e., $P_1=71.5$ s, $P_2=53.7$ s, and $P_3=41.7$ s. They are all close to the periods seen in the case of the p-mode-like driver ($P_1=76.3$ s, $P_2=54.0$ s, and $P_3=41.6$ s). Thus, we are convinced that our interpretation of the harmonic periods illustrated in Figure \ref{fig:vx} and \ref{fig:harmonic} is correct. 

We note that these periods of different harmonics deviate significantly from the results under ideal conditions ($P_1/P_2=2$ and $P_1/P_3=3$). In fact, we have $P_1/P_2=1.33$ and $P_1/P_3=1.72$. Such a deviation can be interpreted as a consequence of a strong density stratification \citep[e.g.,][]{andries2005, erdlyi2007}, and has also been discovered in previous observations \citep[$P_1/P_2=1.4$ obtained in][]{duckenfield2018} and simulations \citep[$P_1/P_2=1.7-1.8$ obtained in][]{afa2020,kara2020}. Our model has a much shorter loop length and includes the lower atmosphere, so a larger deviation is expected.

\begin{figure}
    \centering
    \includegraphics[width=0.48\textwidth]{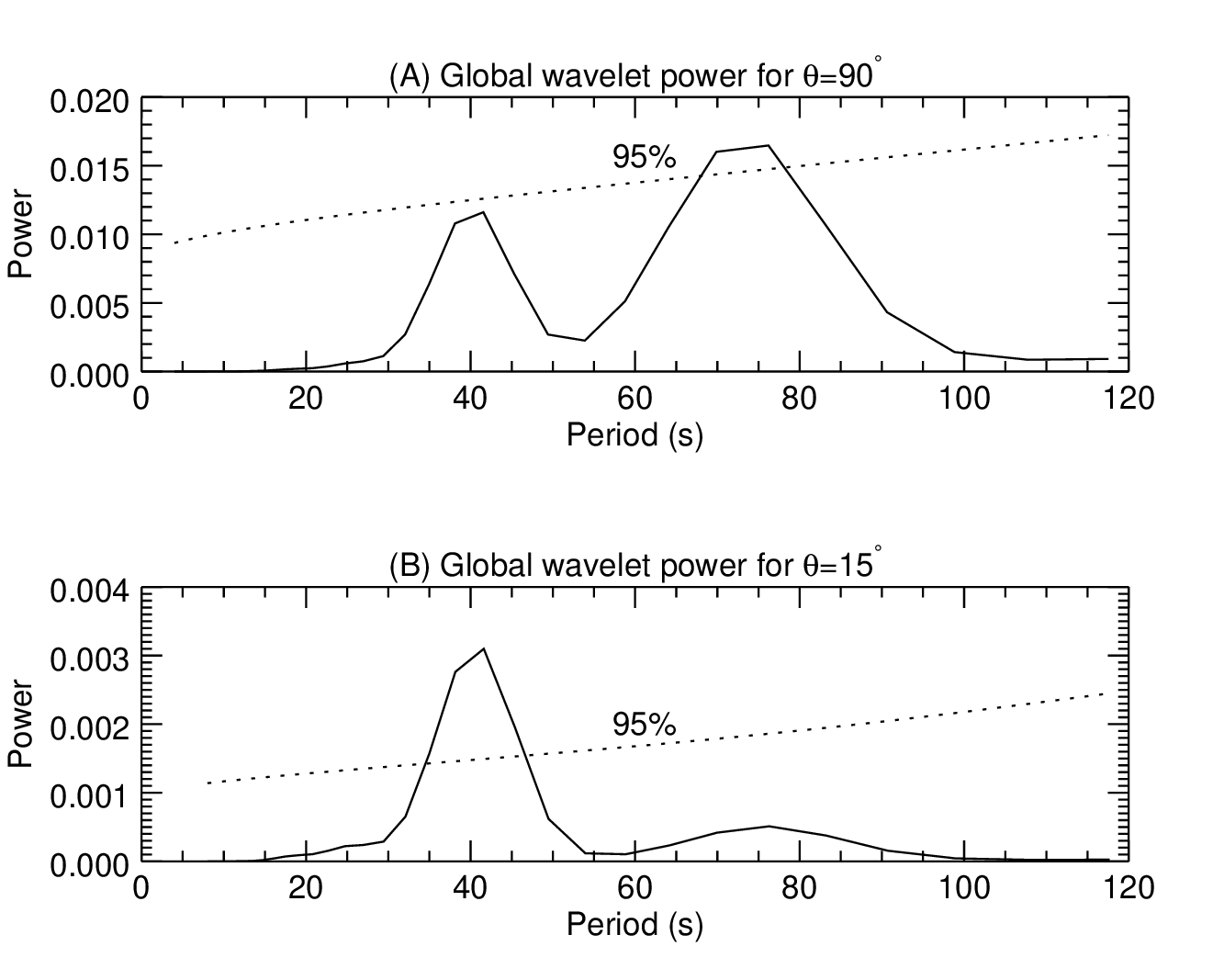}
    \caption{Global wavelet spectrum for a period range of 0-120 s for cases of different inclination angles ($\theta$). (A) $\theta=90^\circ$. (B) $\theta=15^\circ$. A 95\% significance level is also indicated in both panels with dashed lines.}
    \label{fig:global_ws_compare}
\end{figure}

From Figure \ref{fig:vx}(C) and (D), we can see one thing that is unusual: the power of the third harmonic is considerably higher than that of the fundamental mode. It is not only inconsistent with the traditional harmonic wave theory where more power in the fundamental mode is expected, but also different from the results in previous studies \citep{duckenfield2018,afa2020,kara2020}. The exact cause of this result is unknown, but it might be associated with the longitudinal waves also produced by the driver. To demonstrate this, we run another simulation with only the driver's inclination angle $\theta$ changing from 15$^\circ$ to 90$^\circ$. In other words, now the driver is fully horizontal, and longitudinal perturbations are not excited in the loop. In this case, we make a plot similar to Figure \ref{fig:vx}(D) and compare them in Figure \ref{fig:global_ws_compare}. It is shown that for the case where $\theta=90^\circ$, the fundamental mode has a larger power than the third harmonic, which is expected. Therefore, perhaps the longitudinal motions introduced by p-modes can influence the power of higher harmonics. However, the specific reason remains to be investigated in the future, which is beyond the scope of this study.

\subsection{Propagation and cut-off of the transverse motions}\label{subsec:cutoff}

\begin{figure*}
    \centering
    \includegraphics[width=1.0\textwidth]{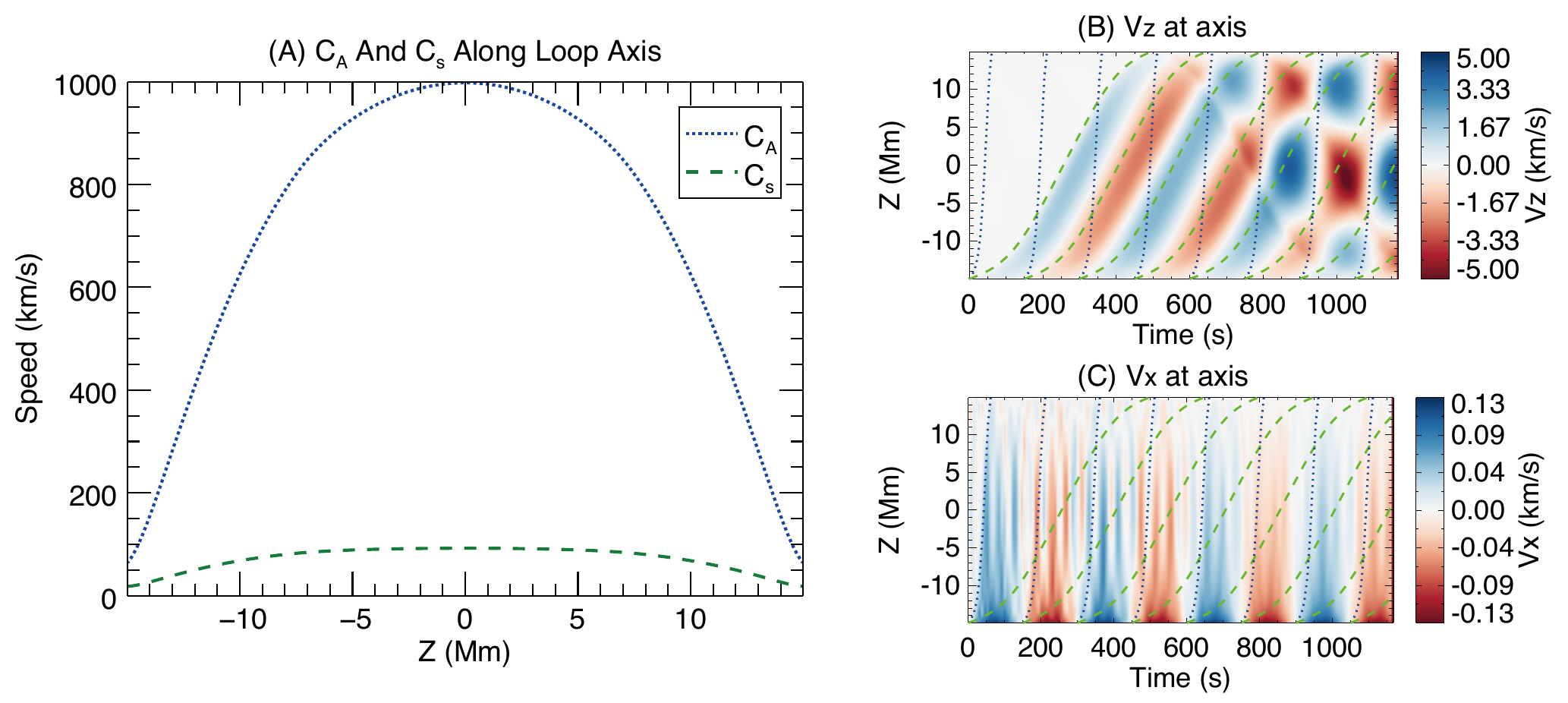}
    \caption{(A) The Alfv\'{e}n speed $c_\mathrm{A}$ and sound speed $c_\mathrm{s}$ along the loop axis. (B-C): Similar to panel (A) of Figure \ref{fig:vz} and Figure \ref{fig:vx}, but we use blue dotted lines and green dashed lines to show the Alfv\'{e}n speed and sound speed, respectively.}
    \label{fig:speeds}
\end{figure*}

In Figure \ref{fig:speeds}(A), we plot the Alfv\'{e}n speed $c_\text{A}$ and sound speed $c_\text{s}$ along the loop axis. The Alfv\'{e}n speed reaches its peak value of about $1000 \text{ km s}^{-1}$ at the height of the loop apex, much larger than the sound speed ($\lesssim90 \text{ km s}^{-1}$).
In Figure \ref{fig:speeds} (B) and (C), some curves corresponding to $c_\text{A}$ (blue dotted lines) and $c_\text{s}$ (green dashed lines) are overplotted in space-time plots of $v_z$ and $v_x$. The slope of these curves corresponds to the local Alfv\'{e}n/sound speed. The propagation property for the excited longitudinal wave can be clearly seen in Figure \ref{fig:speeds}(B), where the curves corresponding to the sound speed have a good consistency with the longitudinal perturbations. It is as expected because the longitudinal wave is interpreted as the slow mode which has a phase speed close to $c_\text{s}$. When it comes to the propagation of transverse motions ($v_x$), we have a more complicated pattern due to the coexistence of short-period eigenmodes and long-period driven oscillations. Nevertheless, we can notice that the establishment of transverse oscillations in the whole loop is much faster than the sound speed. Some large-amplitude $v_x$ signals also have a good correspondence with the Alfv\'{e}n speed. It is easy to understand since the transverse motions driven directly by the driver will travel along the loop with a speed close to $c_\text{A}$ \citep[e.g.,][]{magyar2018}.

The stratification in the transition region could influence the propagation of transverse motions along the loop. 
In \cite{pelouze2023}, the existence of the cut-off of kink waves has been confirmed. They claimed that kink waves with periods larger than the cut-off can reach the corona but with significant damping in amplitude. Although the physical parameters of our shorter loop are different, a similar scenario with a cut-off of about 100 s is still expected. In Figure \ref{fig:vx}(A), we can see the decrease in transverse velocity $v_x$ along the loop. The attenuation is especially significant in the lower atmosphere, as shown in Figure \ref{fig:kink}(A). Nevertheless, the kink waves can still be analysed as aforementioned. To further confirm the existence of the cut-off, we conduct a new run with the same setup but a different driver that has a frequency higher than the cut-off. In this case, the driven kink waves are not subject to the cut-off effect. The result shows that the amplitude increases with height and reaches its peak at $z=0$, rather than experiencing any attenuation. 
Thus, the long-period transverse motion in our simulation is indeed damped by the cut-off, while the standing kink waves with short periods seem to be less affected. 

One limitation of this model is that the broadened transition region would reduce the reflection of kink waves compared to the real atmosphere. As revealed in \cite{Magyar2019}, the vertical gradient in Alfv\'{e}n speed determines the rate of reflection. A steeper transition region may thus prevent the upward propagating waves towards the corona. The observed presence of a 5-min peak in the coronal wave power spectrum \citep[e.g.,][]{tomczyk2009,morton2019,gao2022} could be explained by a number of alternative mechanisms: (1) mode conversion \citep[e.g.,][]{cally2008,khomenko2012}, in which the leakage rate of the 5-min p-mode is not determined by the gradient in Alfv\'{e}n speed; (2) reduced kink mode cut-offs in regions of inclined magnetic field \citep[similar to slow waves,][]{dePontieu2005}; (3) wave-guiding effects in the surroundings of structures (Skirvin et al. 2023, in prep). The influence of these mechanisms on the wave driving requires further investigation.


\section{Summary}\label{sec:summary}

In this study, we modelled a short coronal loop with two footpoints located in the lower atmosphere. 
An inclined p-mode-like driver is employed at one footpoint to excite transverse oscillations in the loop. It is found that two types of decayless oscillations can be excited: the first has a period of about 5 min close to the period of the driver, and the second has shorter periods of several tens of seconds.
We interpreted the former as a propagating wave directly driven by the p-modes, while the latter as kink eigenmodes of different harmonics. Our results are consistent with recent observations of decayless oscillations in short coronal loops, including the high-frequency ones found with SolO/EUI data \citep{zhong2022eui,li2023,Petrova2023} and relatively low-frequency ones found inside CBPs with SDO/AIA \citep{gao2022}. 

The photospheric p-modes have been previously believed to play a key role in the generation of propagating kink waves in the corona \citep{tvd2008,tomczyk2009,morton2015,morton2016,morton2019}, and our findings here suggest a possible relationship between these p-modes and decayless oscillations widely distributed in closed coronal magnetic structures.
If a considerable amount of the energy from p-modes can be transported to the corona, it could contribute to balancing the radiative loss and accelerate the solar wind. 
Given this, the energy analysis in this scenario will be the focus of our future work. Additionally, the decayless transverse oscillation has been seen as a useful seismological tool to diagnose physical properties in the corona \citep[e.g.,][]{tian2012,nistico2013,anfi2019,li2023}. In this regard, our results suggest that one needs to be careful when performing such coronal seismology, because some oscillations may not be kink eigenmodes but driven motions. In that case, the standard seismological scheme is not valid.

Finally, it is of great interest to observationally detect the coexistence of these two types of decayless oscillations in the future. Despite some limitations (as discussed in Section \ref{subsec:2types}), with the unprecedented high spatial and temporal resolution of SolO/EUI, it is still promising to observe such a phenomenon and thus further verify our conclusions.

\acknowledgments

We thank the anonymous referee for the helpful comments that have greatly improved our manuscript.
This work is supported by National Key R\&D Program of China No. 2021YFA1600500 and China Scholarship Council under file No. 202206010018. TVD was supported by the European Research Council (ERC) under the European Union's Horizon 2020 research and innovation programme (grant agreement No 724326), the C1 grant TRACEspace of Internal Funds KU Leuven, and a Senior Research Project (G088021N) of the FWO Vlaanderen. The research benefitted greatly from discussions at ISSI. 

\bibliography{ref}{}

\begin{thebibliography}{}
\expandafter\ifx\csname natexlab\endcsname\relax\def\natexlab#1{#1}\fi
\providecommand{\url}[1]{\href{#1}{#1}}
\providecommand{\dodoi}[1]{doi:~\href{http://doi.org/#1}{\nolinkurl{#1}}}
\providecommand{\doeprint}[1]{\href{http://ascl.net/#1}{\nolinkurl{http://ascl.net/#1}}}
\providecommand{\doarXiv}[1]{\href{https://arxiv.org/abs/#1}{\nolinkurl{https://arxiv.org/abs/#1}}}

\bibitem[{{Afanasyev} {et~al.}(2020){Afanasyev}, {Van Doorsselaere}, \&
  {Nakariakov}}]{afa2020}
{Afanasyev}, A.~N., {Van Doorsselaere}, T., \& {Nakariakov}, V.~M. 2020, \aap,
  633, L8, \dodoi{10.1051/0004-6361/201937187}

\bibitem[{{Andries} {et~al.}(2005){Andries}, {Arregui}, \&
  {Goossens}}]{andries2005}
{Andries}, J., {Arregui}, I., \& {Goossens}, M. 2005, \apjl, 624, L57,
  \dodoi{10.1086/430347}

\bibitem[{{Anfinogentov} {et~al.}(2013){Anfinogentov}, {Nistic{\`o}}, \&
  {Nakariakov}}]{anfin2013}
{Anfinogentov}, S., {Nistic{\`o}}, G., \& {Nakariakov}, V.~M. 2013, \aap, 560,
  A107, \dodoi{10.1051/0004-6361/201322094}

\bibitem[{{Anfinogentov} \& {Nakariakov}(2019)}]{anfi2019}
{Anfinogentov}, S.~A., \& {Nakariakov}, V.~M. 2019, \apjl, 884, L40,
  \dodoi{10.3847/2041-8213/ab4792}

\bibitem[{{Anfinogentov} {et~al.}(2015){Anfinogentov}, {Nakariakov}, \&
  {Nistic{\`o}}}]{anfin2015}
{Anfinogentov}, S.~A., {Nakariakov}, V.~M., \& {Nistic{\`o}}, G. 2015, \aap,
  583, A136, \dodoi{10.1051/0004-6361/201526195}

\bibitem[{{Antolin} {et~al.}(2016){Antolin}, {De Moortel}, {Van Doorsselaere},
  \& {Yokoyama}}]{antolin2016}
{Antolin}, P., {De Moortel}, I., {Van Doorsselaere}, T., \& {Yokoyama}, T.
  2016, \apjl, 830, L22, \dodoi{10.3847/2041-8205/830/2/L22}

\bibitem[{{Aschwanden} {et~al.}(1999){Aschwanden}, {Fletcher}, {Schrijver}, \&
  {Alexander}}]{aschwanden1999}
{Aschwanden}, M.~J., {Fletcher}, L., {Schrijver}, C.~J., \& {Alexander}, D.
  1999, \apj, 520, 880, \dodoi{10.1086/307502}

\bibitem[{{Aschwanden} \& {Schrijver}(2002)}]{aschwanden2002}
{Aschwanden}, M.~J., \& {Schrijver}, C.~J. 2002, \apjs, 142, 269,
  \dodoi{10.1086/341945}

\bibitem[{{Cally}(2017)}]{cally2017}
{Cally}, P.~S. 2017, \mnras, 466, 413, \dodoi{10.1093/mnras/stw3215}

\bibitem[{{Cally} \& {Goossens}(2008)}]{cally2008}
{Cally}, P.~S., \& {Goossens}, M. 2008, \solphys, 251, 251,
  \dodoi{10.1007/s11207-007-9086-3}

\bibitem[{{Chen} \& {Peter}(2015)}]{chen2015}
{Chen}, F., \& {Peter}, H. 2015, \aap, 581, A137,
  \dodoi{10.1051/0004-6361/201526237}

\bibitem[{{De Pontieu} {et~al.}(2005){De Pontieu}, {Erd{\'e}lyi}, \& {De
  Moortel}}]{dePontieu2005}
{De Pontieu}, B., {Erd{\'e}lyi}, R., \& {De Moortel}, I. 2005, \apjl, 624, L61,
  \dodoi{10.1086/430345}

\bibitem[{{Dedner} {et~al.}(2002){Dedner}, {Kemm}, {Kr{\"o}ner}, {Munz},
  {Schnitzer}, \& {Wesenberg}}]{dedner2002}
{Dedner}, A., {Kemm}, F., {Kr{\"o}ner}, D., {et~al.} 2002, Journal of
  Computational Physics, 175, 645, \dodoi{10.1006/jcph.2001.6961}

\bibitem[{{Duckenfield} {et~al.}(2018){Duckenfield}, {Anfinogentov}, {Pascoe},
  \& {Nakariakov}}]{duckenfield2018}
{Duckenfield}, T., {Anfinogentov}, S.~A., {Pascoe}, D.~J., \& {Nakariakov},
  V.~M. 2018, \apjl, 854, L5, \dodoi{10.3847/2041-8213/aaaaeb}

\bibitem[{{Edwin} \& {Roberts}(1983)}]{er1983}
{Edwin}, P.~M., \& {Roberts}, B. 1983, \solphys, 88, 179,
  \dodoi{10.1007/BF00196186}

\bibitem[{{Erd{\'e}lyi} \& {Verth}(2007)}]{erdlyi2007}
{Erd{\'e}lyi}, R., \& {Verth}, G. 2007, \aap, 462, 743,
  \dodoi{10.1051/0004-6361:20065693}

\bibitem[{{Gao} {et~al.}(2022{\natexlab{a}}){Gao}, {Tian}, {Van Doorsselaere},
  \& {Chen}}]{gao2022}
{Gao}, Y., {Tian}, H., {Van Doorsselaere}, T., \& {Chen}, Y.
  2022{\natexlab{a}}, \apj, 930, 55, \dodoi{10.3847/1538-4357/ac62cf}

\bibitem[{{Gao} {et~al.}(2022{\natexlab{b}}){Gao}, {Tian}, {Guo}, \&
  {Li}}]{gao2022review}
{Gao}, Y.~H., {Tian}, H., {Guo}, M.~Z., \& {Li}, B. 2022{\natexlab{b}}, Acta
  Astronomica Sinica, 63, 1

\bibitem[{{Goddard} {et~al.}(2016){Goddard}, {Nistic{\`o}}, {Nakariakov}, \&
  {Zimovets}}]{goddard2016}
{Goddard}, C.~R., {Nistic{\`o}}, G., {Nakariakov}, V.~M., \& {Zimovets}, I.~V.
  2016, \aap, 585, A137, \dodoi{10.1051/0004-6361/201527341}

\bibitem[{{Goossens} {et~al.}(2011){Goossens}, {Erd{\'e}lyi}, \&
  {Ruderman}}]{goossens2011}
{Goossens}, M., {Erd{\'e}lyi}, R., \& {Ruderman}, M.~S. 2011, \ssr, 158, 289,
  \dodoi{10.1007/s11214-010-9702-7}

\bibitem[{{Goossens} {et~al.}(2014){Goossens}, {Soler}, {Terradas}, {Van
  Doorsselaere}, \& {Verth}}]{goossens2014}
{Goossens}, M., {Soler}, R., {Terradas}, J., {Van Doorsselaere}, T., \&
  {Verth}, G. 2014, \apj, 788, 9, \dodoi{10.1088/0004-637X/788/1/9}

\bibitem[{{Guo} {et~al.}(2023){Guo}, {Duckenfield}, {Van Doorsselaere},
  {Karampelas}, {Pelouze}, \& {Gao}}]{guo2023}
{Guo}, M., {Duckenfield}, T., {Van Doorsselaere}, T., {et~al.} 2023, \apjl,
  949, L1, \dodoi{10.3847/2041-8213/acd347}

\bibitem[{{Guo} {et~al.}(2020){Guo}, {Li}, \& {Van Doorsselaere}}]{guo2020}
{Guo}, M., {Li}, B., \& {Van Doorsselaere}, T. 2020, \apj, 904, 116,
  \dodoi{10.3847/1538-4357/abc1df}

\bibitem[{{Guo} {et~al.}(2019{\natexlab{a}}){Guo}, {Van Doorsselaere},
  {Karampelas}, \& {Li}}]{guo2019a}
{Guo}, M., {Van Doorsselaere}, T., {Karampelas}, K., \& {Li}, B.
  2019{\natexlab{a}}, \apj, 883, 20, \dodoi{10.3847/1538-4357/ab338e}

\bibitem[{{Guo} {et~al.}(2019{\natexlab{b}}){Guo}, {Van Doorsselaere},
  {Karampelas}, {Li}, {Antolin}, \& {De Moortel}}]{guo2019b}
{Guo}, M., {Van Doorsselaere}, T., {Karampelas}, K., {et~al.}
  2019{\natexlab{b}}, \apj, 870, 55, \dodoi{10.3847/1538-4357/aaf1d0}

\bibitem[{{Heyvaerts} \& {Priest}(1983)}]{heyvaerts1983}
{Heyvaerts}, J., \& {Priest}, E.~R. 1983, \aap, 117, 220

\bibitem[{{Howson} \& {De Moortel}(2023)}]{howson2023}
{Howson}, T., \& {De Moortel}, I. 2023, Physics, 5, 140,
  \dodoi{10.3390/physics5010011}

\bibitem[{{Howson} \& {De Moortel}(2022)}]{howson2022}
{Howson}, T.~A., \& {De Moortel}, I. 2022, \aap, 661, A144,
  \dodoi{10.1051/0004-6361/202142872}

\bibitem[{{Karampelas} \& {Van Doorsselaere}(2020)}]{kara2020}
{Karampelas}, K., \& {Van Doorsselaere}, T. 2020, \apjl, 897, L35,
  \dodoi{10.3847/2041-8213/ab9f38}

\bibitem[{{Karampelas} \& {Van Doorsselaere}(2021)}]{kara2021}
---. 2021, \apjl, 908, L7, \dodoi{10.3847/2041-8213/abdc2b}

\bibitem[{{Karampelas} {et~al.}(2019{\natexlab{a}}){Karampelas}, {Van
  Doorsselaere}, \& {Guo}}]{kara2019a}
{Karampelas}, K., {Van Doorsselaere}, T., \& {Guo}, M. 2019{\natexlab{a}},
  \aap, 623, A53, \dodoi{10.1051/0004-6361/201834309}

\bibitem[{{Karampelas} {et~al.}(2019{\natexlab{b}}){Karampelas}, {Van
  Doorsselaere}, {Pascoe}, {Guo}, \& {Antolin}}]{kara2019b}
{Karampelas}, K., {Van Doorsselaere}, T., {Pascoe}, D.~J., {Guo}, M., \&
  {Antolin}, P. 2019{\natexlab{b}}, Frontiers in Astronomy and Space Sciences,
  6, 38, \dodoi{10.3389/fspas.2019.00038}

\bibitem[{{Khomenko} \& {Cally}(2012)}]{khomenko2012}
{Khomenko}, E., \& {Cally}, P.~S. 2012, \apj, 746, 68,
  \dodoi{10.1088/0004-637X/746/1/68}

\bibitem[{{Kohutova} \& {Verwichte}(2017)}]{kohutova2017}
{Kohutova}, P., \& {Verwichte}, E. 2017, \aap, 602, A23,
  \dodoi{10.1051/0004-6361/201629912}

\bibitem[{{Lemen} {et~al.}(2012){Lemen}, {Title}, {Akin}, {Boerner}, {Chou},
  {Drake}, {Duncan}, {Edwards}, {Friedlaender}, {Heyman}, {Hurlburt}, {Katz},
  {Kushner}, {Levay}, {Lindgren}, {Mathur}, {McFeaters}, {Mitchell}, {Rehse},
  {Schrijver}, {Springer}, {Stern}, {Tarbell}, {Wuelser}, {Wolfson}, {Yanari},
  {Bookbinder}, {Cheimets}, {Caldwell}, {Deluca}, {Gates}, {Golub}, {Park},
  {Podgorski}, {Bush}, {Scherrer}, {Gummin}, {Smith}, {Auker}, {Jerram},
  {Pool}, {Soufli}, {Windt}, {Beardsley}, {Clapp}, {Lang}, \&
  {Waltham}}]{lemen2012}
{Lemen}, J.~R., {Title}, A.~M., {Akin}, D.~J., {et~al.} 2012, \solphys, 275,
  17, \dodoi{10.1007/s11207-011-9776-8}

\bibitem[{{Li} {et~al.}(2023){Li}, {Bai}, {Tian}, {Su}, {Hou}, {Deng}, {Ji}, \&
  {Ning}}]{li2023sutri}
{Li}, D., {Bai}, X., {Tian}, H., {et~al.} 2023, \aap, 675, A169,
  \dodoi{10.1051/0004-6361/202245812}

\bibitem[{{Li} \& {Long}(2023)}]{li2023}
{Li}, D., \& {Long}, D.~M. 2023, \apj, 944, 8, \dodoi{10.3847/1538-4357/acacf4}

\bibitem[{{Lionello} {et~al.}(2009){Lionello}, {Linker}, \&
  {Miki{\'c}}}]{lionello2009}
{Lionello}, R., {Linker}, J.~A., \& {Miki{\'c}}, Z. 2009, \apj, 690, 902,
  \dodoi{10.1088/0004-637X/690/1/902}

\bibitem[{{Madjarska}(2019)}]{madjarska2019}
{Madjarska}, M.~S. 2019, Living Reviews in Solar Physics, 16, 2,
  \dodoi{10.1007/s41116-019-0018-8}

\bibitem[{{Magyar} \& {Van Doorsselaere}(2018)}]{magyar2018}
{Magyar}, N., \& {Van Doorsselaere}, T. 2018, \apj, 856, 144,
  \dodoi{10.3847/1538-4357/aab42c}

\bibitem[{{Magyar} {et~al.}(2019){Magyar}, {Van Doorsselaere}, \&
  {Goossens}}]{Magyar2019}
{Magyar}, N., {Van Doorsselaere}, T., \& {Goossens}, M. 2019, \apj, 882, 50,
  \dodoi{10.3847/1538-4357/ab357c}

\bibitem[{{Mandal} {et~al.}(2021){Mandal}, {Tian}, \& {Peter}}]{mandal2021}
{Mandal}, S., {Tian}, H., \& {Peter}, H. 2021, \aap, 652, L3,
  \dodoi{10.1051/0004-6361/202141542}

\bibitem[{{Mignone} {et~al.}(2007){Mignone}, {Bodo}, {Massaglia}, {Matsakos},
  {Tesileanu}, {Zanni}, \& {Ferrari}}]{mignone2007}
{Mignone}, A., {Bodo}, G., {Massaglia}, S., {et~al.} 2007, \apjs, 170, 228,
  \dodoi{10.1086/513316}

\bibitem[{{Miki{\'c}} {et~al.}(2013){Miki{\'c}}, {Lionello}, {Mok}, {Linker},
  \& {Winebarger}}]{mikic2013}
{Miki{\'c}}, Z., {Lionello}, R., {Mok}, Y., {Linker}, J.~A., \& {Winebarger},
  A.~R. 2013, \apj, 773, 94, \dodoi{10.1088/0004-637X/773/2/94}

\bibitem[{{Morton} {et~al.}(2015){Morton}, {Tomczyk}, \& {Pinto}}]{morton2015}
{Morton}, R.~J., {Tomczyk}, S., \& {Pinto}, R. 2015, Nature Communications, 6,
  7813, \dodoi{10.1038/ncomms8813}

\bibitem[{{Morton} {et~al.}(2016){Morton}, {Tomczyk}, \& {Pinto}}]{morton2016}
{Morton}, R.~J., {Tomczyk}, S., \& {Pinto}, R.~F. 2016, \apj, 828, 89,
  \dodoi{10.3847/0004-637X/828/2/89}

\bibitem[{{Morton} {et~al.}(2019){Morton}, {Weberg}, \&
  {McLaughlin}}]{morton2019}
{Morton}, R.~J., {Weberg}, M.~J., \& {McLaughlin}, J.~A. 2019, Nature
  Astronomy, 3, 223, \dodoi{10.1038/s41550-018-0668-9}

\bibitem[{{M{\"u}ller} {et~al.}(2020){M{\"u}ller}, {St. Cyr}, {Zouganelis},
  {Gilbert}, {Marsden}, {Nieves-Chinchilla}, {Antonucci}, {Auch{\`e}re},
  {Berghmans}, {Horbury}, {Howard}, {Krucker}, {Maksimovic}, {Owen}, {Rochus},
  {Rodriguez-Pacheco}, {Romoli}, {Solanki}, {Bruno}, {Carlsson}, {Fludra},
  {Harra}, {Hassler}, {Livi}, {Louarn}, {Peter}, {Sch{\"u}hle}, {Teriaca}, {del
  Toro Iniesta}, {Wimmer-Schweingruber}, {Marsch}, {Velli}, {De Groof},
  {Walsh}, \& {Williams}}]{muller2020}
{M{\"u}ller}, D., {St. Cyr}, O.~C., {Zouganelis}, I., {et~al.} 2020, \aap, 642,
  A1, \dodoi{10.1051/0004-6361/202038467}

\bibitem[{{Nakariakov} {et~al.}(2016){Nakariakov}, {Anfinogentov},
  {Nistic{\`o}}, \& {Lee}}]{naka2016}
{Nakariakov}, V.~M., {Anfinogentov}, S.~A., {Nistic{\`o}}, G., \& {Lee}, D.~H.
  2016, \aap, 591, L5, \dodoi{10.1051/0004-6361/201628850}

\bibitem[{{Nakariakov} \& {Ofman}(2001)}]{naka2001}
{Nakariakov}, V.~M., \& {Ofman}, L. 2001, \aap, 372, L53,
  \dodoi{10.1051/0004-6361:20010607}

\bibitem[{{Nakariakov} {et~al.}(1999){Nakariakov}, {Ofman}, {Deluca},
  {Roberts}, \& {Davila}}]{naka1999}
{Nakariakov}, V.~M., {Ofman}, L., {Deluca}, E.~E., {Roberts}, B., \& {Davila},
  J.~M. 1999, Science, 285, 862, \dodoi{10.1126/science.285.5429.862}

\bibitem[{{Nakariakov} {et~al.}(2021){Nakariakov}, {Anfinogentov}, {Antolin},
  {Jain}, {Kolotkov}, {Kupriyanova}, {Li}, {Magyar}, {Nistic{\`o}}, {Pascoe},
  {Srivastava}, {Terradas}, {Vasheghani Farahani}, {Verth}, {Yuan}, \&
  {Zimovets}}]{naka2021}
{Nakariakov}, V.~M., {Anfinogentov}, S.~A., {Antolin}, P., {et~al.} 2021, \ssr,
  217, 73, \dodoi{10.1007/s11214-021-00847-2}

\bibitem[{{Nistic{\`o}} {et~al.}(2013){Nistic{\`o}}, {Nakariakov}, \&
  {Verwichte}}]{nistico2013}
{Nistic{\`o}}, G., {Nakariakov}, V.~M., \& {Verwichte}, E. 2013, \aap, 552,
  A57, \dodoi{10.1051/0004-6361/201220676}

\bibitem[{{Pagano} \& {De Moortel}(2019)}]{pagano2019}
{Pagano}, P., \& {De Moortel}, I. 2019, \aap, 623, A37,
  \dodoi{10.1051/0004-6361/201834158}

\bibitem[{{Pelouze} {et~al.}(2023){Pelouze}, {Van Doorsselaere}, {Karampelas},
  {Riedl}, \& {Duckenfield}}]{pelouze2023}
{Pelouze}, G., {Van Doorsselaere}, T., {Karampelas}, K., {Riedl}, J.~M., \&
  {Duckenfield}, T. 2023, arXiv e-prints, arXiv:2301.03100,
  \dodoi{10.48550/arXiv.2301.03100}

\bibitem[{{Pesnell} {et~al.}(2012){Pesnell}, {Thompson}, \&
  {Chamberlin}}]{pesnell2012}
{Pesnell}, W.~D., {Thompson}, B.~J., \& {Chamberlin}, P.~C. 2012, \solphys,
  275, 3, \dodoi{10.1007/s11207-011-9841-3}

\bibitem[{{Petrova} {et~al.}(2023){Petrova}, {Magyar}, {Van Doorsselaere}, \&
  {Berghmans}}]{Petrova2023}
{Petrova}, E., {Magyar}, N., {Van Doorsselaere}, T., \& {Berghmans}, D. 2023,
  \apj, 946, 36, \dodoi{10.3847/1538-4357/acb26a}

\bibitem[{{Poedts} {et~al.}(1990){Poedts}, {Goossens}, \&
  {Kerner}}]{poedts1990}
{Poedts}, S., {Goossens}, M., \& {Kerner}, W. 1990, \apj, 360, 279,
  \dodoi{10.1086/169118}

\bibitem[{{Riedl} {et~al.}(2021){Riedl}, {Van Doorsselaere}, {Reale},
  {Goossens}, {Petralia}, \& {Pagano}}]{riedl2021}
{Riedl}, J.~M., {Van Doorsselaere}, T., {Reale}, F., {et~al.} 2021, \apj, 922,
  225, \dodoi{10.3847/1538-4357/ac23c7}

\bibitem[{{Riedl} {et~al.}(2019){Riedl}, {Van Doorsselaere}, \&
  {Santamaria}}]{riedl2019}
{Riedl}, J.~M., {Van Doorsselaere}, T., \& {Santamaria}, I.~C. 2019, \aap, 625,
  A144, \dodoi{10.1051/0004-6361/201935393}

\bibitem[{{Rochus} {et~al.}(2020){Rochus}, {Auch{\`e}re}, {Berghmans}, {Harra},
  {Schmutz}, {Sch{\"u}hle}, {Addison}, {Appourchaux}, {Aznar Cuadrado},
  {Baker}, {Barbay}, {Bates}, {BenMoussa}, {Bergmann}, {Beurthe}, {Borgo},
  {Bonte}, {Bouzit}, {Bradley}, {B{\"u}chel}, {Buchlin}, {B{\"u}chner},
  {Cab{\'e}}, {Cadiergues}, {Chaigneau}, {Chares}, {Choque Cortez}, {Coker},
  {Condamin}, {Coumar}, {Curdt}, {Cutler}, {Davies}, {Davison}, {Defise}, {Del
  Zanna}, {Delmotte}, {Delouille}, {Dolla}, {Dumesnil}, {D{\"u}rig}, {Enge},
  {Fran{\c{c}}ois}, {Fourmond}, {Gillis}, {Giordanengo}, {Gissot}, {Green},
  {Guerreiro}, {Guilbaud}, {Gyo}, {Haberreiter}, {Hafiz}, {Hailey}, {Halain},
  {Hansotte}, {Hecquet}, {Heerlein}, {Hellin}, {Hemsley}, {Hermans}, {Hervier},
  {Hochedez}, {Houbrechts}, {Ihsan}, {Jacques}, {J{\'e}r{\^o}me}, {Jones},
  {Kahle}, {Kennedy}, {Klaproth}, {Kolleck}, {Koller}, {Kotsialos},
  {Kraaikamp}, {Langer}, {Lawrenson}, {Le Clech'}, {Lenaerts}, {Liebecq},
  {Linder}, {Long}, {Mampaey}, {Markiewicz-Innes}, {Marquet}, {Marsch},
  {Matthews}, {Mazy}, {Mazzoli}, {Meining}, {Meltchakov}, {Mercier}, {Meyer},
  {Monecke}, {Monfort}, {Morinaud}, {Moron}, {Mountney}, {M{\"u}ller},
  {Nicula}, {Parenti}, {Peter}, {Pfiffner}, {Philippon}, {Phillips},
  {Plesseria}, {Pylyser}, {Rabecki}, {Ravet-Krill}, {Rebellato}, {Renotte},
  {Rodriguez}, {Roose}, {Rosin}, {Rossi}, {Roth}, {Rouesnel}, {Roulliay},
  {Rousseau}, {Ruane}, {Scanlan}, {Schlatter}, {Seaton}, {Silliman}, {Smit},
  {Smith}, {Solanki}, {Spescha}, {Spencer}, {Stegen}, {Stockman}, {Szwec},
  {Tamiatto}, {Tandy}, {Teriaca}, {Theobald}, {Tychon}, {van Driel-Gesztelyi},
  {Verbeeck}, {Vial}, {Werner}, {West}, {Westwood}, {Wiegelmann}, {Willis},
  {Winter}, {Zerr}, {Zhang}, \& {Zhukov}}]{rochus2020}
{Rochus}, P., {Auch{\`e}re}, F., {Berghmans}, D., {et~al.} 2020, \aap, 642, A8,
  \dodoi{10.1051/0004-6361/201936663}

\bibitem[{{Shi} {et~al.}(2021){Shi}, {Van Doorsselaere}, {Guo}, {Karampelas},
  {Li}, \& {Antolin}}]{shi2021}
{Shi}, M., {Van Doorsselaere}, T., {Guo}, M., {et~al.} 2021, \apj, 908, 233,
  \dodoi{10.3847/1538-4357/abda54}

\bibitem[{{Skirvin} {et~al.}(2023){Skirvin}, {Gao}, \& {Van
  Doorsselaere}}]{skirvin2023}
{Skirvin}, S., {Gao}, Y., \& {Van Doorsselaere}, T. 2023, arXiv e-prints,
  arXiv:2304.01606, \dodoi{10.48550/arXiv.2304.01606}

\bibitem[{{Skirvin} {et~al.}(2022){Skirvin}, {Fedun}, {Silva}, \&
  {Verth}}]{skirvin2022}
{Skirvin}, S.~J., {Fedun}, V., {Silva}, S. S.~A., \& {Verth}, G. 2022, \mnras,
  510, 2689, \dodoi{10.1093/mnras/stab3635}

\bibitem[{{Soler} {et~al.}(2021){Soler}, {Terradas}, {Oliver}, \&
  {Ballester}}]{soler2021}
{Soler}, R., {Terradas}, J., {Oliver}, R., \& {Ballester}, J.~L. 2021, \apj,
  909, 190, \dodoi{10.3847/1538-4357/abdec5}

\bibitem[{{Su} {et~al.}(2018){Su}, {Guo}, {Erd{\'e}lyi}, {Ning}, {Ding},
  {Cheng}, \& {Tan}}]{su2018}
{Su}, W., {Guo}, Y., {Erd{\'e}lyi}, R., {et~al.} 2018, Scientific Reports, 8,
  4471, \dodoi{10.1038/s41598-018-22796-7}

\bibitem[{{Terradas} {et~al.}(2008){Terradas}, {Andries}, {Goossens},
  {Arregui}, {Oliver}, \& {Ballester}}]{terradas2008}
{Terradas}, J., {Andries}, J., {Goossens}, M., {et~al.} 2008, \apjl, 687, L115,
  \dodoi{10.1086/593203}

\bibitem[{{Tian} {et~al.}(2008){Tian}, {Curdt}, {Marsch}, \& {He}}]{tian2008}
{Tian}, H., {Curdt}, W., {Marsch}, E., \& {He}, J. 2008, \apjl, 681, L121,
  \dodoi{10.1086/590410}

\bibitem[{{Tian} {et~al.}(2012){Tian}, {McIntosh}, {Wang}, {Ofman}, {De
  Pontieu}, {Innes}, \& {Peter}}]{tian2012}
{Tian}, H., {McIntosh}, S.~W., {Wang}, T., {et~al.} 2012, \apj, 759, 144,
  \dodoi{10.1088/0004-637X/759/2/144}

\bibitem[{{Tomczyk} \& {McIntosh}(2009)}]{tomczyk2009}
{Tomczyk}, S., \& {McIntosh}, S.~W. 2009, \apj, 697, 1384,
  \dodoi{10.1088/0004-637X/697/2/1384}

\bibitem[{{Van Doorsselaere} {et~al.}(2021){Van Doorsselaere}, {Goossens},
  {Magyar}, {Ruderman}, \& {Ismayilli}}]{tvd2021}
{Van Doorsselaere}, T., {Goossens}, M., {Magyar}, N., {Ruderman}, M.~S., \&
  {Ismayilli}, R. 2021, \apj, 910, 58, \dodoi{10.3847/1538-4357/abe630}

\bibitem[{{Van Doorsselaere} {et~al.}(2007){Van Doorsselaere}, {Nakariakov}, \&
  {Verwichte}}]{tvd2007}
{Van Doorsselaere}, T., {Nakariakov}, V.~M., \& {Verwichte}, E. 2007, \aap,
  473, 959, \dodoi{10.1051/0004-6361:20077783}

\bibitem[{{Van Doorsselaere} {et~al.}(2008){Van Doorsselaere}, {Nakariakov},
  {Young}, \& {Verwichte}}]{tvd2008}
{Van Doorsselaere}, T., {Nakariakov}, V.~M., {Young}, P.~R., \& {Verwichte}, E.
  2008, \aap, 487, L17, \dodoi{10.1051/0004-6361:200810186}

\bibitem[{{Van Doorsselaere} {et~al.}(2020){Van Doorsselaere}, {Srivastava},
  {Antolin}, {Magyar}, {Vasheghani Farahani}, {Tian}, {Kolotkov}, {Ofman},
  {Guo}, {Arregui}, {De Moortel}, \& {Pascoe}}]{tvd2020}
{Van Doorsselaere}, T., {Srivastava}, A.~K., {Antolin}, P., {et~al.} 2020,
  \ssr, 216, 140, \dodoi{10.1007/s11214-020-00770-y}

\bibitem[{{Verwichte} {et~al.}(2013){Verwichte}, {Van Doorsselaere}, {White},
  \& {Antolin}}]{verwichte2013}
{Verwichte}, E., {Van Doorsselaere}, T., {White}, R.~S., \& {Antolin}, P. 2013,
  \aap, 552, A138, \dodoi{10.1051/0004-6361/201220456}

\bibitem[{{Wang} {et~al.}(2007){Wang}, {Innes}, \& {Qiu}}]{Wang2007}
{Wang}, T., {Innes}, D.~E., \& {Qiu}, J. 2007, \apj, 656, 598,
  \dodoi{10.1086/510424}

\bibitem[{{Wang} {et~al.}(2012){Wang}, {Ofman}, {Davila}, \& {Su}}]{wang2012}
{Wang}, T., {Ofman}, L., {Davila}, J.~M., \& {Su}, Y. 2012, \apjl, 751, L27,
  \dodoi{10.1088/2041-8205/751/2/L27}

\bibitem[{{Yang} {et~al.}(2020{\natexlab{a}}){Yang}, {Tian}, {Tomczyk},
  {Morton}, {Bai}, {Samanta}, \& {Chen}}]{yang2020b}
{Yang}, Z., {Tian}, H., {Tomczyk}, S., {et~al.} 2020{\natexlab{a}}, Science in
  China E: Technological Sciences, 63, 2357, \dodoi{10.1007/s11431-020-1706-9}

\bibitem[{{Yang} {et~al.}(2020{\natexlab{b}}){Yang}, {Bethge}, {Tian},
  {Tomczyk}, {Morton}, {Del Zanna}, {McIntosh}, {Karak}, {Gibson}, {Samanta},
  {He}, {Chen}, \& {Wang}}]{yang2020a}
{Yang}, Z., {Bethge}, C., {Tian}, H., {et~al.} 2020{\natexlab{b}}, Science,
  369, 694, \dodoi{10.1126/science.abb4462}

\bibitem[{{Yuan} {et~al.}(2023){Yuan}, {Fu}, {Cao}, {Ku{\'z}ma}, {Geeraerts},
  {Trelles Arjona}, {Murawski}, {Van Doorsselaere}, {Srivastava}, {Miao},
  {Feng}, {Feng}, {Noda}, {Cobo}, \& {Su}}]{Yuan2023}
{Yuan}, D., {Fu}, L., {Cao}, W., {et~al.} 2023, Nature Astronomy, 7, 856,
  \dodoi{10.1038/s41550-023-01973-3}

\bibitem[{{Zhang} {et~al.}(2020){Zhang}, {Dai}, {Xu}, {Li}, {Lu}, {Tam}, \&
  {Xu}}]{zhang2020}
{Zhang}, Q.~M., {Dai}, J., {Xu}, Z., {et~al.} 2020, \aap, 638, A32,
  \dodoi{10.1051/0004-6361/202038233}

\bibitem[{{Zhong} {et~al.}(2022{\natexlab{a}}){Zhong}, {Nakariakov},
  {Kolotkov}, \& {Anfinogentov}}]{zhong2022long}
{Zhong}, S., {Nakariakov}, V.~M., {Kolotkov}, D.~Y., \& {Anfinogentov}, S.~A.
  2022{\natexlab{a}}, \mnras, 513, 1834, \dodoi{10.1093/mnras/stac1014}

\bibitem[{{Zhong} {et~al.}(2022{\natexlab{b}}){Zhong}, {Nakariakov},
  {Kolotkov}, {Verbeeck}, \& {Berghmans}}]{zhong2022eui}
{Zhong}, S., {Nakariakov}, V.~M., {Kolotkov}, D.~Y., {Verbeeck}, C., \&
  {Berghmans}, D. 2022{\natexlab{b}}, \mnras, 516, 5989,
  \dodoi{10.1093/mnras/stac2545}

\bibitem[{{Zimovets} \& {Nakariakov}(2015)}]{zimovets2015}
{Zimovets}, I.~V., \& {Nakariakov}, V.~M. 2015, \aap, 577, A4,
  \dodoi{10.1051/0004-6361/201424960}

\end{thebibliography}
\bibliographystyle{aasjournal}

\end{CJK*}
\end{document}